\documentclass[11pt,a4paper,fleqn]{scrartcl}

\usepackage{natbib}
\bibpunct[]{[}{]}{,}{}{}{,}

\usepackage[dvips]{color,graphicx}
\DeclareGraphicsExtensions{.eps}

\newtheorem{definition}{Definition}
\newtheorem{property}{Property}

\title{Reciprocal best hits are not a logically sufficient condition for
  orthology}

\author{Toby Johnson ${}^{1,*}$}

\date{\today}

\begin{document}

\bibliographystyle{toby}

\maketitle

\begin{enumerate}
\item D\'epartement de G\'en\'etique M\'edicale\\
  Universit\'e de Lausanne, CH
 \item[*] 
   email: \texttt{toby.johnson@unil.ch}
\end{enumerate}


\begin{abstract}
  \begin{center}
    \textbf{Summary}
  \end{center}
  It is common to use reciprocal best hits, also known as a
  boom\-er\-ang criterion, for determining orthology between
  sequences.  The best hits may be found by \textsc{blast}, or by
  other more recently developed algorithms.  Previous work seems to
  have assumed that reciprocal best hits is a sufficient but not
  necessary condition for orthology.  In this article, I explain why
  reciprocal best hits cannot logically be a sufficient condition for
  orthology.  If reciprocal best hits is neither sufficient nor
  necessary for orthology, it would seem worthwhile to examine further
  the logical foundations of some unsupervised algorithms that are
  used to identify orthologs.
\end{abstract}


\clearpage
\section{Introduction}
\label{sec:introduction}
\emph{Orthology} between sequences means that they are vertically
descended from a single common ancestral sequence \citep[see
e.g.][]{koonin05:_orthol_paral_evolut_genom}.  Many studies of molecular
evolution rely on comparison of orthologous sequences from different
species.  With increasing amounts of sequence data available,
increasing use is being made of unsupervised algorithms to identify
orthologous sequences.  A reciprocal best \textsc{blast} hit condition
has been widely used to identify orthologs.  Although this approach
can be refined \cite[e.g.][]{wall03:_detec_putat_orthol}, the idea of
a reciprocal best hit remains central to many methods for ortholog
detection.

A reciprocal best hits method makes use of an algorithm, such as
\textsc{blast}, that allows a \emph{query sequence} to be queried
against a \emph{database} of sequences, and returns a ranked list of
\emph{hits}, which are sequences in the database that are similar to
the query sequence.  The top ranked hit is the \emph{best hit}.
Consider the task of identifying orthologs, when many sequences from
each of two species are available.  There are therefore two databases
of sequences, with one database for each species.  Then:
\begin{definition}
  \label{def:pairwise-rbbh}
  Sequences $s_1$ and $s_2$, in databases $G_1$ and $G_2$
  respectively, are said to be (pairwise) reciprocal best
  hits if:
  \begin{itemize}
  \item[(i)] $s_2$ is the best hit when $s_1$ is queried against $G_2$, and
  \item[(ii)] $s_1$ is the best hit when $s_2$ is queried against $G_1$.
  \end{itemize}
\end{definition}

When $s_1$ and $s_2$ are reciprocal best hits, it is common to assume
that they are orthologs.  This is sometimes called the boomerang
condition for (assumption of) orthology.  The metaphor is that, if the
condition is satisfied, one can start at $s_1$ (in $G_1$), go to
$s_1$'s best hit in $G_2$, which is $s_2$, and then go to $s_2$'s best
hit in $G_1$, and end up back where one started.

The notion of reciprocal best hits extends to three or
more species.  A natural definition is:
\begin{definition}
  \label{def:nway-rbbh}
  Sequences $s_1$, $s_2$, \ldots, $s_m$, in databases $G_1$, $G_2$,
  \ldots, $G_m$ respectively, are said to be ($m$-way) reciprocal best
   hits if:
  \begin{itemize}
  \item[(i)] $s_j$ is the best hit when $s_i$ is queried
    against $G_j$
  \end{itemize}
  for all $i,j\in\{1,2,\ldots,m\}$.
\end{definition}

If $s_1$, $s_2$, \ldots, $s_m$ are $m$-way reciprocal best hits, it would seem natural to assume that they are a
set of $m$ orthologs.  When we have sequences from $n$ species, some
historically orthologous sequences may have been lost in some present day
species.  Therefore, we presumably do not wish to restrict ourselves
to sets of $m=n$ orthologs, but will also be interested in finding
sets of $m<n$ orthologs.

\cite{wall03:_detec_putat_orthol} say that their reciprocal smallest
distance method finds more sets of orthologs than reciprocal best
\textsc{blast} hits, and that the sets of orthologs found by their
method is a superset of those found by reciprocal best \textsc{blast}
hits.  This means that they consider reciprocal best \textsc{blast}
hits to be a sufficient but not a necessary condition.
\cite{poptsova07:_branc_clust_phylog_algor_selec_gene_famil} state
that the ``reciprocal [best] \textsc{blast} hit method is very
stringent and succeeds in the selection of conserved orthologs with a
low false positive rate, but it often fails to assemble sets of
orthologs in the presence of paralogs''.  This means that they also
consider it to be a sufficient (with probability close to one) but not
a necessary condition.

A key property of orthology, which stems directly from its biological
definition, is that it is transitive.  That is:
\begin{property}[Transitive orthology]
\label{prop:transitive}
  If sequences $s_1$ and
$s_2$ are orthologous, and $s_2$ and $s_3$ are orthologous, then $s_1$
and $s_3$ must also be orthologous.
\end{property}
Note also that by definition:
\begin{property}
\label{prop:samespecies}
 Two different sequences from the same species cannot be orthologous.
\end{property}

Most readers of this journal will be very familiar with everything I
have said above.  The purpose of this article is to point out the
slightly surprising fact that definition~\ref{def:nway-rbbh} above,
for reciprocal best hits, \emph{logically cannot} be a sufficient
condition for orthology.  It contradicts
properties~\ref{prop:transitive} and \ref{prop:samespecies}.  This
result does not depend on any technical details of the algorithm used
to find or rank the hits.

\section{Theoretical Results}
\label{sec:theoretical-results}
First, it is useful to note that definition~\ref{def:nway-rbbh} above,
is equivalent to the following:
\begin{definition}
  \label{def:n-pw-rbbh}
  Sequences $s_1$, $s_2$, \ldots, $s_m$, in databases $G_1$, $G_2$,
  \ldots, $G_m$ respectively, are said to be ($m$-way) reciprocal best
   hits if:
  \begin{itemize}
  \item[(i)] $s_i$ and $s_j$ are pairwise reciprocal best
     hits, according to
    definition~\ref{def:pairwise-rbbh}
  \end{itemize}
  for all pairs $i\not=j\in\{1,2,\ldots,m\}$.
\end{definition}
The exact equivalence between definition~\ref{def:nway-rbbh} and
definition~\ref{def:n-pw-rbbh} can be seen, by observing that
definition~\ref{def:nway-rbbh} will be satisfied if
definition~\ref{def:n-pw-rbbh} is satisfied, and \textit{vice versa}.
This equivalence is perhaps obvious, but it is worth emphasizing that
definition~\ref{def:nway-rbbh} can be rewritten in terms of only
pairwise relationships between sequences.

The problem with definition~\ref{def:nway-rbbh} (and therefore also
with definition~\ref{def:n-pw-rbbh}) can be explained concisely using
a few terms from graph theory.  In graph theory, a \emph{graph} is an
object that consists of \emph{vertices} (or points), and \emph{edges}
(or lines) that connect some of the vertices.  A \emph{clique} is a
part of the graph (i.e.\ a \emph{subgraph}) for which there is an edge
between every pair of vertices.  We will mostly be interested in
cliques that are not subgraphs of larger cliques, which are
technically known as \emph{maximal cliques}.

\begin{figure}[tbp]
  \begin{center}
    \includegraphics[width=\textwidth]{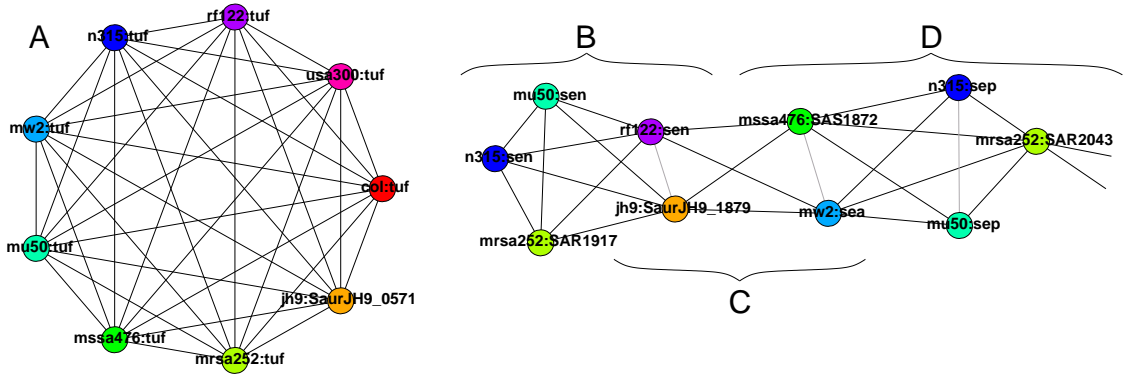}
  \end{center}
  \caption{Subgraph of a reciprocal best \textsc{blastp} hit graph.
    Vertices represent coding sequences, are labelled by
    \textsf{species:gene}, and are colored according to species.
    Edges are drawn between vertices that are pairwise reciprocal best
    \textsc{blastp} hits.  This subgraph contains one nonoverlapping
    maximal clique (A), and three distinct and overlapping maximal
    cliques (B--D).}
  \label{fig:example}
\end{figure}

Consider a graph where every sequence, in every species, is
represented by a vertex.  Let there be an edge between two vertices
if-and-only-if those two edges are pairwise reciprocal best hits.
Then (using definition~\ref{def:n-pw-rbbh}) sets of sequences that are
($m$-way) reciprocal best hits are the cliques of this graph.  A well
known property of cliques is that it is possible for a single vertex
to be a member of more than one maximal clique.  The graph in
figure~\ref{fig:example}
illustrates this property: it contains two five-member maximal cliques
(labelled B and D), each of which has two sequences in common with a
four-member maximal clique (labelled C).  (None of the cliques B--D is
a subgraph of any other, so they are three distinct maximal cliques.)

The example in figure~\ref{fig:example} clearly illustrates why an
$m$-way reciprocal best hit cannot be a sufficient condition for
orthology.  If all the members of clique B are orthologs, and also all
the members of clique C are orthologs, then
property~\ref{prop:transitive} (transitive orthology) means that the
set of sequences, belonging to either clique B or clique C, must all
be orthologs.  The same argument can be extended to clique D.  We
would be logically forced to conclude that, if reciprocal best hits is
sufficient for orthology, then the set of sequences, belonging to any
of cliques B--D, must all be orthologs.  This contradicts the simple
property~\ref{prop:samespecies} of orthologs, because it would mean
that e.g.\ two sequences from the same species mu50, \textit{sep} and
\textit{sen}, are orthologs.  The fact that this \textit{reductio ad
  absurdum} is possible, means that an $m$-way reciprocal best hit
cannot be a logically sufficient condition for orthology.

The structure of the reciprocal best hits graph means that there
cannot be an edge between two sequences from the same species, but
there is no guarantee that longer range structures in the graph will
be consistent with the assumption that cliques represent sets of
orthologs.  A graph representing true biological orthology, consistent
with properties~\ref{prop:transitive} and \ref{prop:samespecies},
consists only of completely separate cliques, each containing at most
one sequence from each species, and with no edges connecting them to
any sequence not in the clique.  It might then seem that we can avoid
the logical problem just described, by excluding cliques that have
edges connecting them to any sequence not in the clique.
Unfortunately, this cannot be a logically sufficient condition for
orthology either.  To see this, we can attempt to construct a
meaningful definition:
\begin{definition}[Perfect reciprocal best hits]
  \label{def:perf-rbbh}
  Sequences $s_1$, $s_2$, \ldots, $s_m$, in databases $G_1$, $G_2$,
  \ldots, $G_m$ respectively, are said to be perfect ($m$-way)
  reciprocal best hits (with respect to other databases $G_{m+1}$,
  $G_{m+2}$, \ldots) if:
  \begin{itemize}
  \item[(i)] $s_i$ and $s_j$ are pairwise reciprocal best
     hits, according to
    definition~\ref{def:pairwise-rbbh}
  \end{itemize}
  for all pairs $i\not=j\in\{1,2,\ldots,m\}$, and
  \begin{itemize}
  \item[(ii)] $s_i$ and $s_\mathrm{o}$ are not pairwise reciprocal best
     hits, according to
    definition~\ref{def:pairwise-rbbh}
  \end{itemize}
  for all $i\in\{1,2,\ldots,m\}$ and for any $s_\mathrm{o}$ in any
  other database $G_{m+1}$, $G_{m+2}$, \ldots.
\end{definition}

In graph theoretic terminology, sets of sequences that satisfy
definition~\ref{def:perf-rbbh} are both (i) maximal cliques, and also
(ii) \emph{maximal connected subgraphs}.  

Definition~\ref{def:perf-rbbh} cannot be a logically sufficient
condition for orthology either, because it depends on the choice of
``other databases'', without which we cannot construct a meaningful
definition of perfect reciprocal best hits.  For example, consider
again the subgraph in figure~\ref{fig:example}.  If sequences from
only three strains mw2, n315 and mu50 were analysed, the three
sequences mw2:sea, n315:sep and mu50:sep would be perfect reciprocal
best hits.  However, if sequences from rf122 were also analysed, the
same three sequences would not be perfect reciprocal best hits.
Because the set of species actually analysed is determined by
arbitrary choice and convenience, and because orthology is a
biological property that is independent of which species are analysed,
then perfect reciprocal best hits (definition~\ref{def:perf-rbbh})
cannot be a logically sufficient condition for orthology.

\section{Empirical Results}
\label{sec:empirical-results}
The mere fact that structures like the one in figure~\ref{fig:example}
are possible, means that technically speaking, reciprocal best hits
cannot be a logically sufficient condition for orthology.  However, it
might be that this is rarely a practical issue.  I therefore studied
as an example, all coding sequences in nine strains of the bacterium
\textit{Staphylococcus aureus}.  Whether these strains are considered
to be distinct species is irrelevant to the argument being presented
here, so I will use the terms species and strain interchangeably.  I
used all annotated coding sequences, and identified reciprocal best
hits using \textsc{blastp} (i.e.\ querying protein sequence against
protein sequence).  These species are closely related, and so
\textsc{blastp} is expected to be a reliable method for finding best
hits.

\begin{figure}[p]
  \includegraphics[width=\textwidth]{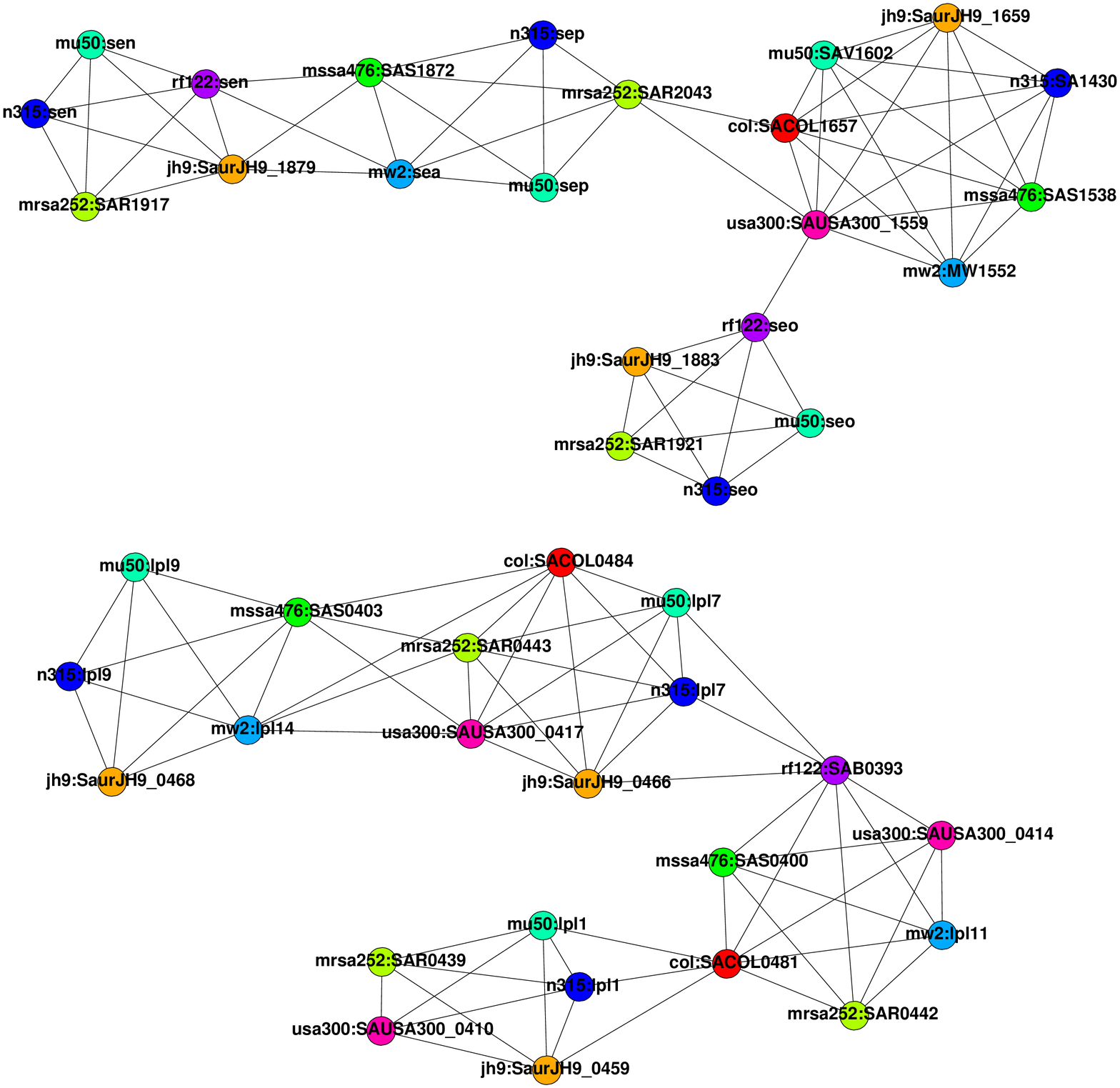}
  \caption{The two largest maximal connected subgraphs, each of size
    22, of the reciprocal best \textsc{blastp} hit graph for nine
    \textit{S. aureus} strains.  Not surprisingly, these graphs
    contain an overrepresentation of coding sequences that have not
    been completely annotated.}
  \label{fig:big-subgraphs}
\end{figure}

The full graph, including all reciprocal best \textsc{blastp} hits,
consists of 3411 maximal connected subgraphs.  That is, there are 3411
smaller graphs, none of which are connected to each other.  Of these,
3282 (96.2\%) are also cliques, and therefore correspond to sets of
sequences that are perfect reciprocal best hits.  The size
distribution of maximal connected subgraphs is given in
table~\ref{tab:mcs-size-dist}.  Figure~\ref{fig:big-subgraphs} shows
the two largest subgraphs, each of size 22.  Each can be seen to
contain several cliques.  Although only 2.8\% of maximal connected
subgraphs are not cliques, these include more of the larger subgraphs.
In total, 5.9\% of all sequences are in maximal connected subgraphs
that are not cliques.  Thus, we cannot determine orthology for over
5\% of sequences using reciprocal best hits, and this proportion would
have to (weakly) increase if more data from more species were
included.

We can also quantify the extent of the problem by finding all cliques
in the reciprocal best hits graph.  There are 3793 cliques in total,
and the size distribution of the cliques is given in
table~\ref{tab:clique-size-dist}.  Considering only cliques of size
three or greater, 13.7\% of cliques are not maximal connected
subgraphs.  This means that for 13.7\% of sets of reciprocal best hit
sequences, we should be cautious about inferring orthology, because
some but not all sequences are reciprocal best hits with other
sequences outside the set.

Thus, even in a closely related group of species/strains, there are
reasonably common problems with using reciprocal best hits to
determine orthology.  It seems likely that the problems will be even
more common with more distantly related species, or when attempting to
identify orthologous regions of noncoding sequence in eukaryotes.

\begin{table}[p]
  \caption{Size distribution of maximal connected subgraphs, for the
    reciprocal best \textsc{blastp} hit graph for nine
    \textit{S. aureus} strains.  The size distribution is broken down
    according to whether the subgraph is a clique or not; subgraphs of
    size greater than nine cannot be cliques.}
  \label{tab:mcs-size-dist}
  \begin{center}
    \begin{tabular}{r r r r r r r r r r r}
      Subgraph size & 1--22 & 1 & 2 & 3 & 4 & 5 & 6 & 7 & 8 & 9 \\
      \hline
      cliques (no.) & 3282 & 581 & 154 & 118 & 77 & 58 & 54 & 72 & 196 & 1972 \\
      (\%) & 96.2 & 17 & 4.5 & 3.5 & 2.3 & 1.7 & 1.6 & 2.1 & 5.7 & 57.8 \\
      (\% sequences) & 94.1 & 2.5 & 1.3 & 1.5 & 1.3 & 1.2 & 1.4 & 2.2 & 6.7 & 76 \\
      \hline
      non-cliques (no.) & 129 & 0 & 0 & 2 & 6 & 7 & 7 & 6 & 11 & 16 \\
      (\%) & 3.8 & 0 & 0 & 0.1 & 0.2 & 0.2 & 0.2 & 0.2 & 0.3 & 0.5 \\
      (\% sequences) & 5.9 & 0 & 0 & 0 & 0.1 & 0.1 & 0.2 & 0.2 & 0.4 & 0.6 \\
      \hline
    \end{tabular}\vspace{16.5pt}

    \begin{tabular}{r r r r r r r r r r r r r r}
      Subgraph size & 10 & 11 & 12 & 13 & 14 & 15 & 16 & 17 & 18 & 19 & 20 & 21 & 22 \\
      \hline
      cliques (no.) & \textsc{na} & \textsc{na} & \textsc{na} & \textsc{na} & \textsc{na} & \textsc{na} & \textsc{na} & \textsc{na} & \textsc{na} & \textsc{na} & \textsc{na} & \textsc{na} & \textsc{na} \\
      \hline
      non-cliques (no.) & 6 & 16 & 15 & 7 & 6 & 5 & 10 & 2 & 2 & 1 & 1 & 1 & 2 \\
      (\%) & 0.2 & 0.5 & 0.4 & 0.2 & 0.2 & 0.1 & 0.3 & 0.1 & 0.1 & 0 & 0 & 0 & 0.1 \\
      (\% sequences) & 0.3 & 0.8 & 0.8 & 0.4 & 0.4 & 0.3 & 0.7 & 0.1 & 0.2 & 0.1 & 0.1 & 0.1 & 0.2 \\
      \hline
    \end{tabular}
  \end{center}
\end{table}
\begin{table}[p]
  \caption{Size distribution of cliques, for the
    reciprocal best \textsc{blastp} hit graph for nine
    \textit{S. aureus} strains.  The size distribution is broken down
    according to whether the clique is also a maximal connected
    subgraph (called ``Perfect'') or not (called ``Imperfect'').
    Percentages are given relative to the total number of cliques of
    size 3 or greater.}
  \label{tab:clique-size-dist}
  \begin{center}
    \begin{tabular}{r r r r r r r r r r r}
      Clique size & 1 & 2 & 3 & 4 & 5 & 6 & 7 & 8 & 9 & all \\
      \hline
      Perfect (no.) & 581 & 154 & 118 & 77 & 58 & 54 & 72 & 196 & 1972 & 3282 \\
      (\%) & \textsc{na} & \textsc{na} & 4 & 2.6 & 2 & 1.8 & 2.4 & 6.6 & 66.8 & 86.3 \\
      \hline
      Imperfect (no.) & 0 & 107 & 94 & 80 & 82 & 75 & 40 & 33 & 0 & 511 \\
      (\%) & \textsc{na} & \textsc{na} & 3.2 & 2.7 & 2.8 & 2.5 & 1.4 & 1.1 & 0 & 13.7 \\
      \hline
      Total & 581 & 261 & 212 & 157 & 140 & 129 & 112 & 229 & 1972 & 3793 \\
      \hline
    \end{tabular}
  \end{center}
\end{table}

\bibliography{tobyrefs}

\end{document}